\documentclass[preprint,aip,jcp]{revtex4-2}
\usepackage{amsmath}
\usepackage{amssymb}
\usepackage{physics}
\usepackage{graphicx} 
\usepackage{enumerate}
\usepackage[inline]{enumitem}
\usepackage{bm}
\usepackage{tcolorbox}
\usepackage{xcolor}
\usepackage[normalem]{ulem}
\usepackage{tikz}

\usepackage{hyperref}
\hypersetup{hidelinks,
  colorlinks=true,
  allcolors=blue,
  pdfstartview=Fit,
  breaklinks=true
}

\newcommand{\im}{\textup{i}}
\newcommand{\Lv}{\mathcal{L}}
\newcommand{\Proj}{\bm{\mathcal{P}}}
\newcommand{\Qroj}{\bm{\mathcal{Q}}}

\begin{document}
\title{Universal Structure of Computing Moments for Exact Quantum Dynamics: Application to Arbitrary System-Bath Couplings}
\author{Rui-Hao Bi}
\affiliation{Department of Chemistry, School of Science and Research Center for Industries of the Future, Westlake University, Hangzhou, Zhejiang 310024, China}

\author{Wei Liu}
\affiliation{Department of Chemistry, School of Science and Research Center for Industries of the Future, Westlake University, Hangzhou, Zhejiang 310024, China}

\author{Wenjie Dou}%
\email{douwenjie@westlake.edu.cn} 
\affiliation{Department of Chemistry, School of Science and Research Center for Industries of the Future, Westlake University, Hangzhou, Zhejiang 310024, China}
\affiliation{Institute of Natural Sciences, Westlake Institute for Advanced Study, Hangzhou, Zhejiang 310024, China}
\affiliation{Key Laboratory for Quantum Materials of Zhejiang Province, Department of Physics, School of Science and Research Center for Industries of the Future, Westlake University, Hangzhou, Zhejiang 310024, China}

\date{\today}

\begin{abstract}
    We introduce a general procedure for computing higher-order moments of correlation functions in open quantum systems, extending the scope of our recent work on Memory Kernel Coupling Theory (MKCT) [\href{https://arxiv.org/abs/2407.01923}{W. Liu, Y. Su, Y. Wang, and W. Dou, arXiv:2407.01923 (2024)}]. This approach is demonstrated for arbitrary system-bath coupling that can be expressed as polynomial, $H_{SB} = \hat{V} (\alpha_0 + \alpha_1 \hat{q} + \alpha_2 \hat{q}^2+ \dots)$, where we show that the recursive commutators of a system operator obey a universal hierarchy. Exploiting this structure, the higher-order moments are obtained by evaluating the expectation values of the system and bath operators separately, with bath expectation values derived from the derivatives of a generating function. We further apply MKCT to compute the dipole autocorrelation function for the spin-boson model with both linear and quadratic coupling, achieving agreement with the hierarchical equations of motion approach. Our findings suggest a promising path toward accurate dynamics for complex open quantum systems. 
\end{abstract}

\maketitle

\section{Introduction}
The two-time correlation function,
\begin{equation}\label{eqn:corr}
C_{AA}(t) = C_{AA}(t, 0) \equiv \Tr\left(e^{\im \hat{H} t}\hat{A} e^{-\im \hat{H} t} \hat{A} \hat{\rho}_0\right),
\end{equation}
captures the quantum dynamics of the operator $\hat{A}$ under the Hamiltonian $\hat{H}$, starting from an initial state $\hat{\rho}_0$. It is fundamental to various experimental observables \cite{harp1970corr} and has recently gained significance in polaritonic chemistry \cite{li2021polaritonic, philbin2022cavity, ke2024cavity}, spectroscopy \cite{tanimura2012twodspec, ma2015spec, saraceno2023spec}, and transport phenomena \cite{li2024transport, jasrasaria2024transport}. In these contexts, the quantum system interacts with a condensed-phase environment consisting of electronic and vibrational degrees of freedom, which can be described by the open quantum system Hamiltonian \cite{Breuer2007OQS}:
\begin{equation}\label{eqn:open_system}
    \hat{H} = \hat{H}_{S} + \hat{H}_B + \hat{H}_{SB},
\end{equation}
where $\hat{H}_S$ represents the system, $\hat{H}_B$ the bath, and $\hat{H}_{SB}$ their interaction.

The significance of $C_{AA}(t)$ in open quantum systems has driven the development of numerous methods, each striking a different balance between accuracy and computational cost.  At the high-accuracy end, exact methods \cite{tanimura1989heom,makri1995quapi,jin2008heom,wang2022DEOM} can solve small model problems but are computationally prohibitive for larger systems. To enable simulations of more complex and realistic models, various approximation methods have been developed, including perturbative approaches \cite{redfield1957redfield,Tokuyama1976TCL}, mode-coupling theory \cite{Reichman2002mct1,Reichman2002mct2}, semi-classical methods \cite{Meyer1979MeyerMiller,Stock1997StockThoss,liu2007lscivr}, and mixed quantum-classical methods \cite{tully90fssh,Craig2005TSHKSDFT,Wang2015SH_progress,Mannouch2023MASH}. Despite these advancements, no single approach has achieved the ideal balance between accuracy and efficiency, and the search for improved methods remains an active area of research.

Projection operator techniques \cite{Nakajima1958projection, zwanzig1960projection, Zwanzig1961projection, mori1965projection} are powerful tools for studying open quantum system dynamics. The Mori projection \cite{mori1965projection}, in particular, simplifies the problem by projecting onto the operator of interest, $\hat{A}$, to reduce the dimensionality. This reformulates the computation of $C_{AA}(t)$ into finding the memory kernel, which offers a key advantage in dissipative systems: it typically decays faster than the correlation function itself \cite{Cohen2011exact_kernel, Dan2022kernel_faster}. However, this approach presents a challenge--the memory kernel depends on the projected propagator \cite{montoya-castillo_approximate_2016}, which is inherently difficult to handle. Significant efforts have been made to compute the memory kernel \cite{Mulvihill2021kernels_rev}, including numerically ``exact" approaches \cite{shi2003memker, Cohen2011exact_kernel, zhang2016deomkernel, Ivander2024unified} and various approximation methods \cite{shi2004semiclassical, kelly_generalized_2016, montoya-castillo_approximate_2016, montoya-castillo_approximate_2017, Bhattacharyya2024mori}.

In a recent study \cite{liu2024mkct}, we introduced Memory Kernel Coupling Theory (MKCT), a novel approach for computing the memory kernel. MKCT defines a set of auxiliary higher-order memory kernels whose evolution is governed solely by the higher-order moments of the correlation function. Since the higher-order moments are pure static quantities, this means MKCT allows us to obtain the correlation function without explicitly evolving the open quantum system. However, our previous implementation relied on the Liouvillian superoperator from Dissipaton Equations of Motion (DEOM) \cite{wang2022DEOM} to compute the moments, limiting the broader applicability of MKCT.

In this article, we present a general procedure for computing correlation function moments to arbitrary order, expanding the applicability of MKCT. We demonstrate our approach using a polynomial bosonic interaction and uncover a universal hierarchy of the recursive commutators of a system operator that determines the moments.

The article is structured as follows: In Sec.~\ref{subsec:mkct}, we revisit the MKCT framework and introduce a new Pad\'{e} approximant-based truncation scheme. Sec.~\ref{subsec:poly} presents our novel higher-order moments computation procedure. In Sec.~\ref{sec:num}, we apply our method to the spin-boson model with linear, quadratic, and combined linear–quadratic coupling. Finally, we conclude in Sec.~\ref{sec:conclusion}.

\section{\label{sec:theory}Theory}
\subsection{Memory Kernel Coupling Theory\label{subsec:mkct}}
Following Mori's projection approach, \cite{mori1965projection} the correlation function in Eq.~\ref{eqn:corr} satisfies the generalized quantum master equation (GQME):  
\begin{equation}
    \dot{C}_{AA}(t) = \Omega C_{AA}(t) + \int_{0}^{t} \dd{\tau} K(t - \tau) C_{AA}(\tau), 
\end{equation}
where the scalar $\Omega$ and the kernel function $K(t)$ are derived from the Mori-Zwanzig formalism:
\begin{equation}
    \Omega = (\im \Lv \hat{A}, A) (\hat{A}, \hat{A})^{-1},
\end{equation}
\begin{equation}
    K(t) = (\im \Lv e^{\im t \Qroj \Lv} \Qroj \im \Lv \hat{A}, \hat{A}) (\hat{A}, \hat{A})^{-1}.
\end{equation}
Here, the Mori-product notation $(\hat{A}, \hat{B})$ denotes an appropriate inner product. The Liouville operator acts on an arbitrary operator $\hat{X}$ as $\Lv \hat{X} \equiv \frac{1}{\hbar} \comm{\hat{H}}{\hat{X}}$. The projection operator is defined as $\Proj \hat{X} = (\hat{X}, \hat{A}) \hat{A} (\hat{A}, \hat{A})^{-1}$, and the complementary projection operator is $\Qroj = \bm{1} - \Proj$.


In practice, $\Omega$ is readily available and many existing works have introduced methods to approximate the kernel function $K(t)$. \cite{kelly_generalized_2016, montoya-castillo_approximate_2016, yan_theoretical_2019} Recently, we have introduced a novel method called Memory Kernel Coupling Theory (MKCT) to compute $K(t)$ efficiently. \cite{liu2024mkct} Our key idea is to extend the definition of $\Omega$ and $K(t)$ in the Mori GQME and introduce the higher-order moments 
\begin{equation}\label{eqn:moments}
    \Omega_n \equiv ((\im\Lv)^n \hat{A}, \hat{A}) (\hat{A}, \hat{A})^{-1},
\end{equation}
and the corresponding auxiliary kernels
\begin{equation}\label{eqn:kernels}
    K_n(t) \equiv ((\im\Lv)^n \hat{f}(t), \hat{A}) (\hat{A}, \hat{A})^{-1},
\end{equation}
where the random fluctuation operator is $\hat{f}(t) = e^{\im t \Qroj\Lv} \Qroj \im\Lv\hat{A}$. 
We have shown in Ref.~\cite{liu2024mkct} that the higher-order kernels satisfy the following coupled ordinary differential equation (ODE): 
\begin{equation}\label{eqn:kn_ode}
    \dot{K}_n(t) = K_{n+1}(t) - \Omega_n K_1(t), 
\end{equation}
with initial conditions $K_{n}(0) = \Omega_{n+1} - \Omega_n \Omega_1$. This result is compelling as it reveals that the quantum dynamics can be fully encoded in a system of interconnected ODEs for the higher-order kernels. Notably, the only parameters needed to integrate $\{K_n(t)\}$ are $\{\Omega_n\}$, meaning that our MKCT framework eliminates the need to explicitly evolve the operator $\hat{A}$ within the open quantum system.

However, a caveat of Eq.~\ref{eqn:kn_ode} is that the equation extends to infinite order, and there is no straightforward method to truncate it at a finite order. From our experience, a hard truncation can lead to numerical instabilities. To this end, we introduce a novel truncation scheme with Pad\'{e} approximant.

To begin with, notice that the $m$-th derivative of kernel $K_n(t)$ evaluated at $t=0$ is 
\begin{equation}\label{eqn:deriv_Kn}
    K_n^{(m)} \equiv ((\im\Lv)^n (\Qroj \im \Lv)^{m+1} \hat{A}, \hat{A}) (\hat{A}, \hat{A})^{-1}.
\end{equation}
A recursion relation can be found by expanding $\Qroj = \bm{1} - \Proj$ that
\begin{equation}\label{eqn:rec}
    K_n^{(m)} = K_{n+1}^{(m-1)} - \Omega_n \tilde{\Omega}_m,
\end{equation}
where we have introduced the auxiliary moment 
\begin{equation}
    \tilde{\Omega}_m = (\im\Lv (\Qroj \im \Lv)^{m} \hat{A}, \hat{A}) (\hat{A}, \hat{A})^{-1}.
\end{equation}
Recursively applying Eq.~\ref{eqn:rec} eventually leads the following expression: 
\begin{equation}
    K_{n}^{(m)} = \Omega_{m+n+1} - \Omega_{n} \tilde{\Omega}_{m} - \dots - \Omega_{m+n} \tilde{\Omega}_{0}, 
\end{equation}
that $K_{n}^{(m)}$ can be expressed with moments and auxiliary moments. Similarly, the auxiliary moments themselves have recursions
\begin{equation}    
\begin{aligned}
    \tilde{\Omega}_m &= (\im\Lv (\Qroj \im \Lv)^{m-1} \im \Lv \hat{A}, \hat{A}) (\hat{A}, \hat{A})^{-1} - \tilde{\Omega}_{m-1} \Omega_1, \\
    &= \Omega_{m+1} - \tilde{\Omega}_{m-1} \Omega_1 - \dots \tilde{\Omega}_0 \Omega_{m}, 
\end{aligned}
\end{equation}
which means the auxiliary moments $\{\tilde{\Omega}_m\}$ can be readily obtained by the moments $\{\Omega_n\}$. 

The series expansion for the $n$-th order kernel can then be given by Pad\'{e} approximant. Initially, $K_n(t)$ is expressed as a truncated Taylor series:
\begin{equation}
    K_n(t) \approx \sum_{j=0}^{M} \frac{K_n^{(j)}(0)}{j!} t^j,
\end{equation}
which is a good local approximation but lacks accuracy over a broader range of $t$. A more reliable approximation can be achieved using the Pad\'{e} approximant: 
\begin{equation}\label{eqn:pade}
    K_n(t) \approx \frac{p_{M_1}(t)}{q_{M_2}(t)} = \frac{a_0 + a_1 t + \dots + a_{M_1} t^{M_1}}{1 + b_1 t + \dots + b_{M_2} t^{M_2}},
\end{equation}
where $ p_{M_1}(t)$ and $q_{M_2}(t)$ are polynomials of degrees $M_1$ and $M_2$, respectively. The coefficients $\{a_i\}$ and $\{b_i\}$ are computed using the python library SciPy, which implements the standard Pad\'{e} approximant procedure as described in Ref.~\cite{Baker1996pade}. Overall, Eq.~\ref{eqn:pade} provides a numerically stable truncation for the MKCT Eq.~\ref{eqn:kn_ode}, where \emph{all} coefficients can be evaluated with higher-order moments $\{ \Omega_n \}$.

\subsection{Moments for a general system-bath model: polynomial interaction\label{subsec:poly}}
To study the dynamics of a discrete quantum system coupled to a vibrational environment, the following model is commonly used: 
\begin{equation}\label{eqn:H_sys_bath_lin}
    \hat{H} = \hat{H}_S + \sum_j \frac{\hbar\omega_j}{2} (\hat{p}_j^2 + \hat{q}_j^2) + \hat{V} U(\hat{q}),
\end{equation}
where ${\hat{q}_j}$ and $\hat{p}_j$ are the position and momentum operators for the $j$-th mode of a harmonic bath with frequency $\omega_j$, and $U(\hat{q})$ represents arbitrary potential energy that can be expressed in polynomial
\begin{equation}\label{eqn:poly}
    U(\hat{q}) = \alpha_0 + \alpha_1 \hat{q} + \alpha_2 \hat{q}^2 + \dots
\end{equation}
The system is coupled to a polynomial of the collective mode, $\hat{q} = \sum_j c_j \hat{q}_j$, which is characterized by the spectral density: 
\begin{equation}\label{eqn:spec_den}
    J(\omega) = \frac{\pi}{2} \sum_j c_j^2 \delta(\omega - \omega_j).
\end{equation} 
Most studies only consider linear coupling, but a growing body of research shows the significance of non-linear interactions \cite{Xu2018Nonlinear,Zhang2020nonlinear,Chen2023Quad,Han2024quad,Bi2024quad}. Here, we consider a general polynomial coupling with coefficients $\{\alpha_i\}$.

Suppose that $\hat{A}$ is the system operator of interest, applying the MKCT to calculate $C_{AA}(t)$ boils down to \begin{enumerate*}
    \item[1)] Compute $(\im\Lv)^n \hat{A}$ to arbitrary order.
    \item[2)] Evaluate the inner product $((\im\Lv)^n \hat{A}, \hat{A})$.
\end{enumerate*} 
In this work, we define the Mori-product as 
\begin{equation}\label{eqn:mori_product}
    (\hat{A}, \hat{B}) = \Tr(\hat{A} \hat{B}^{\dagger} \hat{\sigma}_0 \otimes \hat{\rho}_\text{eq}^{B}), 
\end{equation}
where we assume a factorized initial condition. Here, $\hat{\sigma}_0$  is the initial system density operator, and $\hat{\rho}_\text{eq}^{B} = e^{-\beta \hat{H}_B}/ \Tr(e^{-\beta \hat{H}_B})$ is the equilibrium bath density operator, with $\beta$ being the inverse temperature.

\subsubsection{Application of the Liouvillian \label{subsec:liouvillian}}
To compute $(\im\Lv)^n \hat{A}$ to arbitrary order,  we begin by writing down the first few terms and recognizing the following pattern: 
\begin{equation}\label{eqn:general}
    (\im \Lv)^n \hat{A} = \sum_{\text{all terms}} \hat{O} \otimes \prod_{j} \hat{Q}_{m_{j}} \prod_{k} \hat{P}_{n_k},
\end{equation}
where each term is a product of a system operator $\hat{O}$ and a polynomial of bath operators. Note that we have introduced the following generalized bath mode operators,
\begin{gather}
    \hat{Q}_m = \sum_j c_j \omega_j^{m} \hat{q}_j, \\
    \hat{P}_n = \sum_j c_j \omega_j^{n} \hat{p}_j, 
\end{gather}
to simplify the notation in Eq.~\ref{eqn:general}. The polynomial is organized so that the position operators $\hat{Q}_m$ always appear before the momentum operators $\hat{P}_n$. Note that the collective mode $\hat{q} \equiv \hat{Q}_0$. For brevity, the direct product symbol $\otimes$ is omitted henceforth.

In order to verify that the general pattern of Eq.~\ref{eqn:general} is closed and to outline a practical procedure for computing $(\im\Lv)^n \hat{A}$, we demonstrate below how applying the Liouvillian $\im\Lv = \im\Lv_S + \im\Lv_B +  \im\Lv_{SB}$ to a general term generates a set of new terms. To formalize this, we denote a general term by $\hat{G} = \hat{O} \prod_{j} \hat{Q}_{m_j} \prod_{k} \hat{P}_{n_k}$.

To begin with, the application of $\im\Lv_S$ is straightforward:
\begin{equation}\label{eqn:iLvSG}
    \im\Lv_S \hat{G} = \im \comm{\hat{H}_S}{\hat{O}} \prod_{j} \hat{Q}_{m_j} \prod_{k} \hat{P}_{n_k}.
\end{equation}
Next, we utilize the properties of the harmonic bath outlined in Eqs.~\ref{eqn:iLvBQ} and~\ref{eqn:iLvBP}, along with the commutation relation
$\comm{A}{BC} = \comm{A}{B} C + B\comm{A}{C}$, to write 
\begin{multline}\label{eqn:iLvBG}
    \im \Lv_{B} \hat{G} = \hat{O} \left[\sum_{\alpha} \left(\prod_{j_1 < \alpha} \hat{Q}_{m_{j_1}} \right) \hat{P}_{m_{\alpha}+1}  \left(\prod_{j_2 > \alpha} \hat{Q}_{m_{j_2}} \right)  \right] \prod_k \hat{P}_{n_k} \\ 
    - \hat{O} \prod_{j} \hat{Q}_{m_j}
    \left[\sum_{\alpha} \left(\prod_{k_1 < \alpha} \hat{P}_{n_{\alpha}} \right) \hat{Q}_{m_{\alpha} + 1}  \left(\prod_{k_2 > \alpha} \hat{P}_{m_{k_2}} \right)  \right].
\end{multline}
Note that the terms in Eq.~\ref{eqn:iLvBG} differ from the general form in Eq.~\ref{eqn:general} because the $\hat{Q}_n$ terms do not always appear before the $\hat{P}_n$ terms. To this end, Eq.~\ref{eqn:iLvBG} can be further ordered using the following procedures:
\begin{gather}
    \hat{P}_n \prod_{j} \hat{Q}_{{m}_j} = \prod_{j} \hat{Q}_{{m}_j} \hat{P}_n  + \sum_{\alpha} (-\im \theta_{n+m_{\alpha}}) \prod_{j\neq\alpha} \hat{Q}_{{m}_j}, \label{eqn:order_P_prod_Q}\\
    \prod_{k} \hat{P}_{n_k} \hat{Q}_{m} = \hat{Q}_{m} \prod_{k} \hat{P}_{n_k} + \sum_{\alpha}  (-\im \theta_{m+n_{\alpha}}) \prod_{k\neq\alpha} \hat{P}_{{n}_k}, \label{eqn:order_prod_P_Q}
\end{gather}
where we have used the harmonic bath property from Eq.~\ref{eqn:comm_Q_P} and the identity $\hat{A}\hat{B} = \hat{B}\hat{A} + \comm{\hat{A}}{\hat{B}}$. Note that we have introduced we introduce the function $\theta_n$,
\begin{equation}\label{eqn:theta}
    \theta_n = \frac{2}{\pi} \int_{0}^{\infty} \dd{\omega} J(\omega) \omega^{n}, 
\end{equation}
to simplify the notation. Lastly, we expand the commutator $\im \Lv_{SB}$:
\begin{equation} \label{eqn:iLvSBG}
    \im \Lv_{SB} \hat{G} = \im \hat{V} \hat{O} U(\hat{q}) \prod_{j} \hat{Q}_{m_j} \prod_{k} \hat{P}_{n_k}
    - \im \hat{O} \hat{V} \prod_{j} \hat{Q}_{m_j}  \prod_{k} \hat{P}_{n_k} U(\hat{q}), 
\end{equation}
In the second term in Eq.~\ref{eqn:iLvSBG}, the potential operator $U(\hat{q})$
appears after the momentum operators, which is not ordered in the same way as the general term (Eq.~\ref{eqn:general}). However, this term can be brought into the desired order by recursively applying Eq.~\ref{eqn:order_prod_P_Q}.

Overall, Eqs.~\ref{eqn:iLvSG}-\ref{eqn:order_prod_P_Q} and Eq.~\ref{eqn:iLvSBG} demonstrate that the complete structure of $(\im \Lv)^n \hat{A}$ is captured by the general expression (Eq.~\ref{eqn:general}). This hierarchical structure is universal for a system-bath model with arbitrary electron-phonon coupling potential (Eq.~\ref{eqn:H_sys_bath_lin},~\ref{eqn:poly}), which can be recursively computed up to any arbitrary order $n$.

\subsubsection{Evaluation of the Mori products \label{subsec:expval}}
To evaluate the moments $\Omega_n$ (Eq.~\ref{eqn:moments}), we must compute the Mori product between $(\im\Lv)^n \hat{A}$ and $\hat{A}$. This reduces to summing the Mori products of general terms $\{\hat{G}\}$ and $\hat{A}$, where how to obtain the general terms $\hat{G}$ is described in Sec.~\ref{subsec:liouvillian}. These Mori products naturally decompose into separate system and bath expectation values:
\begin{equation}\label{eqn:gen_mori_product}
    (\hat{G}, \hat{A}) = \Tr(\hat{O} \hat{A} \hat{\sigma}_0)\Tr(\prod_{j} \hat{Q}_{m_j} \prod_{k} \hat{P}_{n_k} \hat{\rho}_\text{eq}^{B}).
\end{equation}
The first trace on the right-hand side (RHS) of Eq.~\ref{eqn:gen_mori_product} is straightforward to evaluate. However, the second trace, involving bath operator polynomials, is somewhat cumbersome to evaluate.

In order to compute the expectation value of a bath polynomial, we define the following generating function $f(\{\lambda_{\alpha}\}, \{\lambda_{\gamma}'\})$:
\begin{equation}
    f(\{\lambda_{\alpha}\}, \{\lambda_{\gamma}'\}) = \Tr[\exp(\sum_{\gamma} \lambda'_{\gamma} \hat{P}_{n_{\gamma}}) \hat{\rho}_\text{eq}^{B} \exp(\sum_{\alpha} \lambda_{\alpha} \hat{Q}_{m_{\alpha}})],\label{eqn:def_gen_func}
\end{equation}
such that the second trace in Eq.~\ref{eqn:gen_mori_product} becomes the derivatives of the generating function $f(\{\lambda_{\alpha}\}, \{\lambda_{\gamma}'\})$
\begin{equation}
    \Tr(\prod_{j} \hat{Q}_{m_j} \prod_{k} \hat{P}_{n_k} \hat{\rho}_\text{eq}^{B}) = \prod_{\alpha} \dv{}{\lambda_{\alpha}} \eval_{\lambda_{\alpha}=0} \prod_{\gamma} \dv{}{\lambda_{\gamma}'} \eval_{\lambda_{\gamma}'=0} f(\{\lambda_{\alpha}\}, \{\lambda_{\gamma}'\}).
\end{equation}
As detailed in Appendix~\ref{app:deriv_gen_func}, the generating function has the closed-form expression
\begin{multline}\label{eqn:expval_gen_func}
    f(\{\lambda_{\alpha}\}, \{\lambda_{\gamma}'\}) = \exp[\sum_{\alpha} \sum_{\alpha'} \frac{\lambda_{\alpha} \lambda_{\alpha'}}{2} \eta_{m_{\alpha} + m_{\alpha'}}] \times \\
    \exp[\sum_{\gamma} \sum_{\gamma'}  \frac{\lambda_{\gamma}' \lambda_{\gamma'}'}{2} \eta_{n_{\gamma} + n_{\gamma'}}] \exp[\im \sum_{\alpha} \sum_{\gamma}  \frac{\lambda_{\alpha} \lambda_{\gamma}'}{2} \theta_{m_{\alpha} + n_{\gamma}}],
\end{multline}
where the function $\eta_n$ is defined as
\begin{equation}\label{eqn:eta}
    \eta_{n} = \frac{2}{\pi} \int_{0}^{\infty} \dd{\omega} J(\omega) \omega^{n} \coth(\frac{\hbar\omega\beta}{2}) .
\end{equation}

\subsection{Workflow and comments\label{subsec:theory_comments}}
\begin{figure}[htbp]
    \centering
    \includegraphics[width=0.4\linewidth]{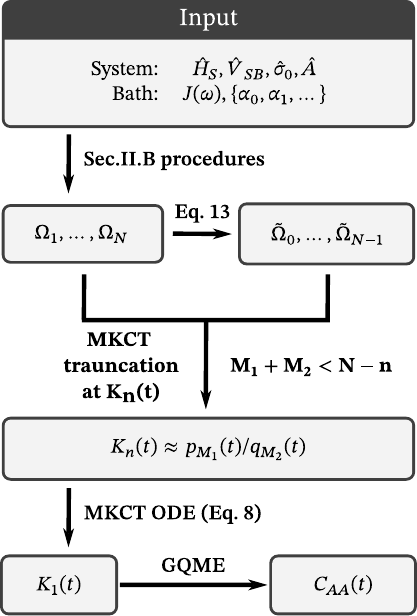}
    \caption{Schematic illustration of the workflow for computing the correlation function using the framework introduced in this work.}
    \label{fig:workflow}
\end{figure}
We summarize the workflow of our novel framework for computing the correlation function in FIG.~\ref{fig:workflow}. Notably, the computation does not require simulating open quantum system dynamics. Instead, the moments $\{\Omega_n\}$ are computed statically under the factorized initial condition. Furthermore, the complexity of the hierarchical structure of $(\im\Lv)^n \hat{A}$ remains independent of the basis size of $\hat{A}$. This suggests that our approach can be applied to larger problems where accurate non-Markovian methods, such as the Hierarchical Equations of Motion, become computationally infeasible.

Note that in the workflow shown in FIG.~\ref{fig:workflow}, the integral functions $\theta_n$ (Eq.\ref{eqn:theta}) and $\eta_n$ (Eq.~\ref{eqn:eta}), which are effectively moment-like integrals over the spectral density $J(\omega)$, must be evaluated. However, these integrals are known to diverge for certain spectral densities, such as the Drude–Lorentz and Brownian oscillator models. Consequently, the current implementation of our method is restricted to spectral densities that yield convergent integrals, such as Ohmic and Gaussian forms.

\section{\label{sec:num}Numerical Example}
In the following, we apply our MKCT framework (Sec.~\ref{subsec:mkct}) 
to compute correlation functions, where the correlation function moments $\Omega_n$ are obtained using the methods introduced in Sec.~\ref{subsec:poly}. Throughout Sec.~\ref{sec:num}, we adopt the Ohmic spectral density
\begin{equation}\label{eqn:exp_spe}
    J(\omega) = 2\lambda\omega e^{-\abs{\omega}/\omega_D}, 
\end{equation}
where $\lambda$ is a coefficient with units of inverse frequency squared, ensuring that $J(\omega)$ has units of inverse frequency. The parameter $\omega_D$ represents the cutoff frequency. In our examples, we set $\hbar = 1$. Note that our particular choice of $J(\omega)$, which has units of inverse frequency, allows us to define the system-bath coupling operator $\hat{V}$ as dimensionless, and ensures that the polynomial coefficients ${\alpha_i}$ carry units of energy, regardless of the order of the polynomial interaction.

We investigate the dipole autocorrelation function, defined as $C_{\mu\mu}(t) = \expval{\hat{\mu}(t)\hat{\mu}(0)}$ where the dipole operator is given by $\hat{\mu} = \hat{\sigma}_x$. The correlation function $C_{\mu\mu}(t)$ plays a crucial role in spectroscopy, as the absorption lineshape function $I(\omega)$ is directly proportional to its Fourier transform \cite{ma2015spec}:
\begin{equation}
    I(\omega) \propto \Re \int_0^{\infty} \dd{t} C_{\mu\mu}(t) e^{\im \omega t}.
\end{equation}

As a reference, we use the Dissipaton Equations of Motion (DEOM) approach \cite{wang2022DEOM} for linear system-bath coupling (Sec.\ref{subsec:sb_linear}) and the extended DEOM approach \cite{Xu2018Nonlinear} for quadratic system-bath coupling (Sec.\ref{subsec:sb_quad}). To efficiently decompose the bath correlation function, we employ the time-domain Prony method \cite{zihao2022prony}.

Lastly, to obtain converged results for the first-order kernel $K_1(t)$ and the correlation function $C_{AA}(t)$. We summarize our choices for the MKCT cutoff and Pad\'{e} approximant parameters, $n$, $M_1$, and $M_2$ (see the workflow in FIG.~\ref{fig:workflow}). While the MKCT hierarchy can, in principle, be truncated at any order, we consistently set $n=1$ for simplicity, as higher-order starting points offered little practical advantage despite faster kernel decay. The kernel $K_1(t)$ is then approximated using Eq.~\ref{eqn:pade} with selected $M_1$ and $M_2$, which in turn determines the required highest moment order: $N<M_1+M_2+n$. Currently, there is no systematic procedure for selecting $M_1$ and $M_2$. Beyond increasing the total Pad\'{e} order $M=M_1 + M_2$, we find that setting $M_1 < M_2$ generally yields better numerical stability.

\subsection{Spin-boson model with linear interaction \label{subsec:sb_linear}}

\begin{figure}[htbp]
    \centering
    \includegraphics[width=0.99\linewidth]{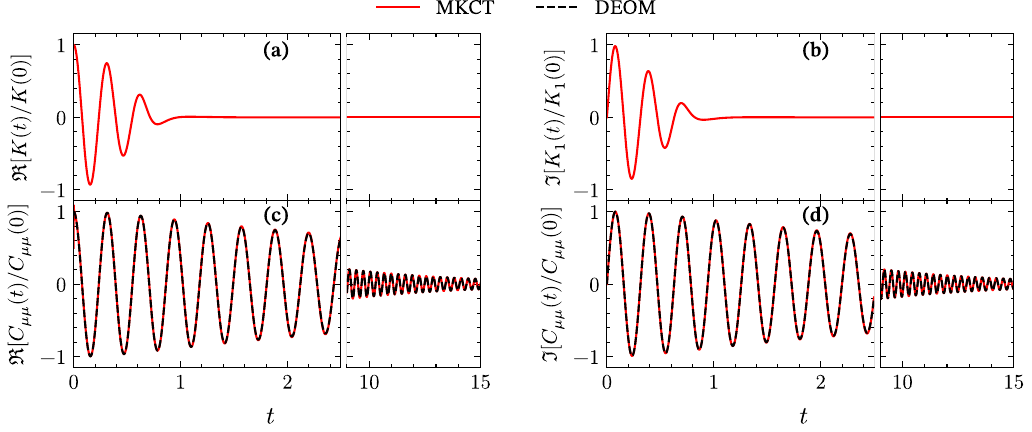}
    \caption{Memory kernel $K_1(t)$ and autocorrelation function $C_{\mu\mu}(t)$ for the spin–boson model with linear interaction. (a-b) are the real and complex part of $K_1(t)$, respectively. (c-d) are the real and complex part of $C_{\mu\mu}(t)$, respectively. The parameters used are $\alpha_0=\alpha_2=0$, $\alpha_1=1$, $\Delta = 20$, $\lambda=0.5$, $\omega_D=1$, and $\beta = 5$. The Pad\'{e} orders $n=1$, $M_1=7$ and $M_2=15$.}
    \label{fig:linear_sb_time}
\end{figure}

\begin{figure}[htbp]
    \centering
    \includegraphics[width=0.49\linewidth]{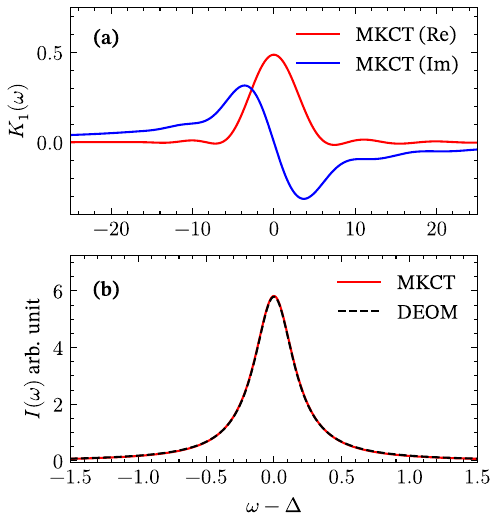}
    \caption{Frequency domain (a) memory kernel $K_1(\omega)$ and (b) absorption lineshape $I(\omega)$ for the spin–boson model with linear interaction. The parameters are the same as those in FIG.~\ref{fig:linear_sb_time}.}
    \label{fig:linear_sb_freq}
\end{figure}

We consider a system Hamiltonian $\hat{H}_S = \frac{\Delta}{2} \sigma_z$, a coupling operator $\hat{V} = \sigma_z$, and an initial state $\hat{\sigma}_0 = \ketbra{0}{0}$. FIG.~\ref{fig:linear_sb_time} presents $K_1(t)$ and $C_{\mu\mu}(t)$ for an energy gap of $\Delta = 20$, demonstrating that our MKCT approach matches the DEOM results exactly. Due to the large energy gap, the correlation function exhibits rapid oscillations, but FIG.~\ref{fig:linear_sb_time} shows that $K_1(t)$ decays to zero much faster than $C_{\mu\mu}(t)$. This highlights the advantage of computing the memory kernel $K_1(t)$, as it captures the dynamics over a shorter timescale. FIG.~\ref{fig:linear_sb_freq} presents the frequency-domain representations $K_1(\omega)$ and $I(\omega)$, showing that the memory kernel $K_1(\omega)$ distribution is much more dispersed than the lineshape function $I(\omega)$. This broader distribution suggests that the memory kernel encodes a wider range of dynamical frequencies, which facilitates efficient modeling of non-Markovian effects. 

In FIG.~\ref{fig:linear_sb_time}, we demonstrate the effectiveness of our method using a biased spin-boson model with fast system dynamics, moderate bath coupling, and low temperature, which are conditions relevant to molecular spectroscopy. For a broader evaluation across different parameter regimes, we refer the reader to Appendix~\ref{app:parameters}, where additional benchmarks and results are provided.

\subsection{Spin-boson model with quadratic interaction \label{subsec:sb_quad}}

Here, we consider a quadratic interaction case adopting the model in Ref.~\cite{Xu2018Nonlinear}. Specifically, we consider a system Hamiltonian $\hat{H}_S = \omega_{eg} \ketbra{1}{1}$, a coupling operator $\hat{V}=\ketbra{1}{1}$, and initial state $\hat{\sigma}_0 = \ketbra{0}{0}$. 

\begin{figure}[htbp]
    \centering
    \includegraphics[width=0.99\linewidth]{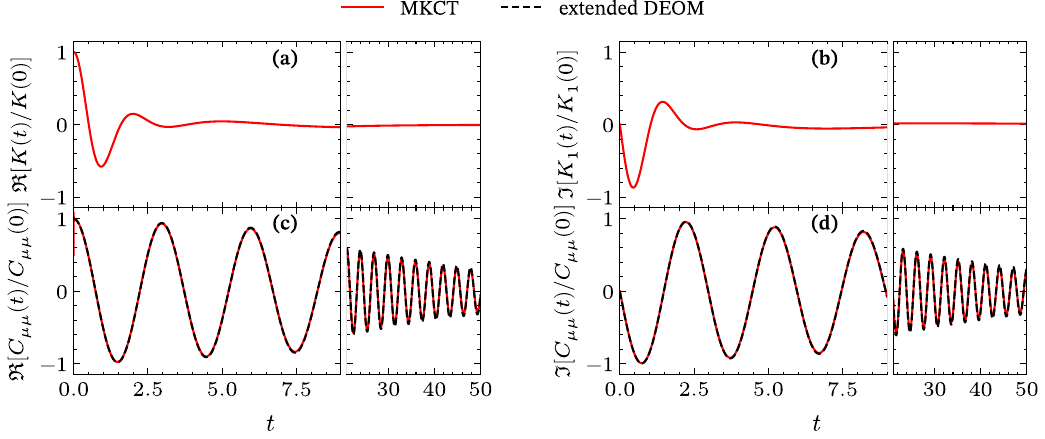}
    \caption{$K_1(t)$ and $C_{\mu\mu}(t)$ of the spin–boson model with purely quadratic interaction. (a-d) follow the panel conventions in FIG.~\ref{fig:linear_sb_time}. The parameters used are  $\alpha_0=\alpha_1=0$, $\alpha_2=0.384$, $\omega_{eg}= 2$, $\lambda=0.5$, $\omega_D=0.5$, and $\beta = 1$. The Pad\'{e} orders $n=1$, $M_1=7$ and $M_2=8$.}
    \label{fig:quad_sb_time}
\end{figure}

\begin{figure}[htbp]
    \centering
    \includegraphics[width=0.99\linewidth]{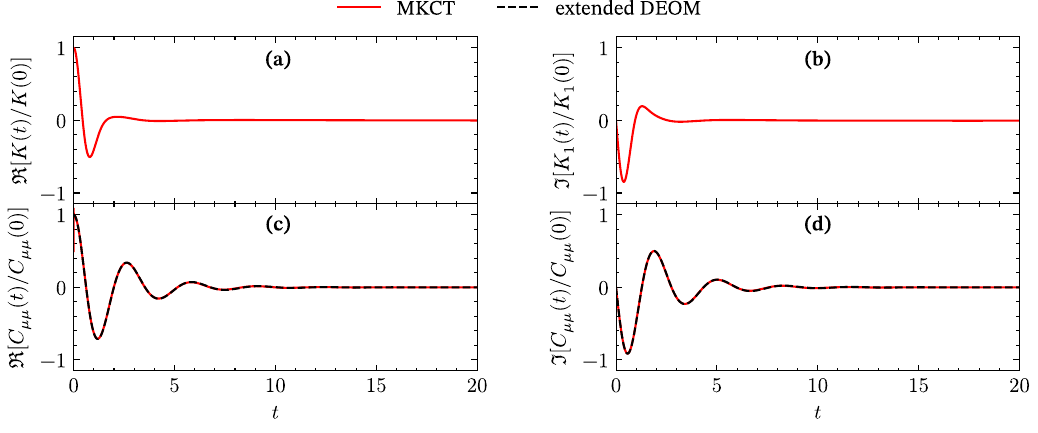}
    \caption{$K_1(t)$ and $C_{\mu\mu}(t)$ of the spin–boson model with both linear and quadratic coupling. (a-d) follow the panel conventions in FIG.~\ref{fig:linear_sb_time}. The parameters used are  $\alpha_0=0.442$, $\alpha_1=-1.251$, $\alpha_2 = 0.384$ , $\omega_{eg}= 2$, $\lambda=0.5$, $\omega_D=0.5$, and $\beta = 1$. The Pad\'{e} orders $n=1$, $M_1=6$ and $M_2=9$.}
    \label{fig:quad_linear_sb_time}
\end{figure}

\begin{figure}[htbp]
    \centering
    \includegraphics[width=0.49\linewidth]{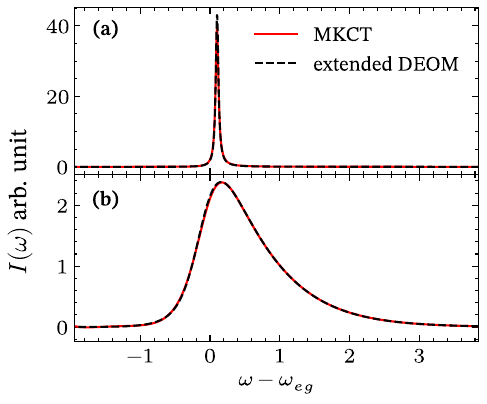}
    \caption{The absorption lineshape $I(\omega)$ of the spin–boson model with quadratic interaction. (a) corresponds to the pure quadratic interaction, with parameters given in FIG.~\ref{fig:quad_sb_time}. (b) represents the mixed linear-quadratic interaction, with parameters given in FIG.~\ref{fig:quad_linear_sb_time}}
    \label{fig:quad_sb_freq}
\end{figure}

FIG.~\ref{fig:quad_sb_time} plots $K_1(t)$ and $C_{\mu\mu}(t)$ for a purely quadratic system-bath interaction. The dynamics induced by $\hat{q}^2$-coupling is known to be highly oscillatory and decay slowly \cite{zihao2023CaldeiraLeggett,Bi2024quad}. However, by leveraging a short-timescale memory kernel, our MKCT approach accurately captures the long-time behavior of $C_{\mu\mu}(t)$. When both linear and quadratic interactions are present, the linear coupling introduces an additional dissipation channel, leading to a faster decay of $C_{\mu\mu}(t)$,  as shown in FIG.~\ref{fig:quad_linear_sb_time}. This effect is correctly captured by our MKCT framework. The difference between purely quadratic dynamics and mixed linear-quadratic dynamics is clearly reflected in the absorption lineshape, with the former exhibiting a more localized peak. As shown in FIG.~\ref{fig:quad_sb_freq}, our MKCT approach accurately captures both the peak position and linewidth.

\section{Conclusions\label{sec:conclusion}}
We have introduced a novel procedure for computing higher-order moments of correlation functions in open quantum systems with polynomial bosonic coupling. By integrating this approach with Memory Kernel Coupling Theory, we efficiently obtain the memory kernel without explicitly evolving the open quantum system. The faster decay of the memory kernel enables the computation of correlation functions over longer timescales. Our approach allows simulation of the spin-boson model with both linear and quadratic coupling, achieving agreement with numerically ``exact" methods.

Our method presents two primary limitations that, if addressed, could expand its applicability. First, the evaluation of $\theta_n$ and $\eta_n$ restricts the method to spectral densities with well-defined moments, such as the Ohmic spectral density. The divergence issue for spectral densities with ill-defined moments, such as the Drude–Lorentz model, might be addressed using regularization techniques. However, our preliminary attempts yielded inconsistent results, and thus they are not included in this work. Second, while the Pad\'{e} approximant-based truncation yields converged results for the kernel, the cutoff parameters $M_1$ and $M_2$ must be empirically tuned, with no systematic procedure currently available for their selection. Developing a more systematic approach to choosing these parameters would improve the robustness of the method.

In this work, we focus on the autocorrelation function with factorized initial conditions. However, our formulation can readily generalize to other types of correlation functions (see Ref.~\cite{liu2024mkct}). Moreover, the Mori-based GQME with factorized initial conditions is formally equivalent to the Zwanzig formulation of the GQME for non-equilibrium dynamics \cite{kelly_generalized_2016,montoya-castillo_approximate_2016}. If one is interested in correlation functions with the true equilibrium density, this framework can be extended to multiple-time correlation functions.

It is also worth noting that the conceptual framework of the method presented in this work is not restricted to bosonic interactions. This flexibility makes it potentially applicable to more complex systems, such as fermionic systems with electron-electron interactions, including Fermi gases and the Hubbard model. These systems pose computational challenges for hierarchical equations of motion, but may be more tractable using our MKCT approach. However, computing moments for such cases is more difficult within the current framework, and modifications to our procedure may be necessary. Work is ongoing to extend and refine these methods for broader applicability.

\begin{acknowledgments}
W.D. acknowledges the support from National Natural Science Foundation of China (No. 22361142829 and No. 22273075) and Zhejiang Provincial Natural Science Foundation (No. XHD24B0301).  R.-H. Bi acknowledges helpful discussion of truncation with Yoshitaka Tanimura. We thank Westlake university supercomputer center for the facility support and technical assistance.
\end{acknowledgments}

\section*{Data Availability}
The data and code implementing the theory presented in Sec.~\ref{sec:theory} are available from the Zenodo repository [Ref.\citenum{bi_2025_15393237}].

\appendix
\section{Properties of harmonic bath\label{app:bath}}
The bare harmonic bath Hamiltonian is given by
\begin{equation}
    \hat{H}_B = \sum_j \frac{\hbar\omega_j}{2}(\hat{p}_j^2 +  \hat{q}_j^2) = \sum_j \hbar \omega_j (\hat{b}_j^\dagger \hat{b}_j + 1/2),
\end{equation}
where the bath mode operators can be represented in ladder operators 
\begin{equation}\label{eqn:ladder_qp}
    \hat{q}_j \equiv \frac{\hat{b}_{j}^{\dagger} + \hat{b}_j}{\sqrt{2}}, \quad \hat{p}_j \equiv \frac{\hat{b}_j - \hat{b}_{j}^{\dagger} }{\im \sqrt{2}}.
\end{equation}
It is straightforward to show 
\begin{equation}\label{eqn:iLvB_components}
    \im \Lv_B \hat{q}_j = \omega_j \hat{p}_j, \quad \im \Lv_B \hat{p}_j = -\omega_j \hat{q}_j. 
\end{equation} 

With properties Eqs.~\ref{eqn:iLvB_components}, we can derive the following results for the generalized bath modes $\hat{Q}_m$ and $\hat{P}_n$:
\begin{gather}
    \im \Lv_B \hat{Q}_m = \hat{P}_{m+1}, \label{eqn:iLvBQ}\\
    \im \Lv_B \hat{P}_m = - \hat{Q}_{m+1}, \label{eqn:iLvBP}\\
    \comm{\hat{Q}_m}{\hat{P}_n} = \frac{2\im}{\pi} \int_{0}^{\infty} \dd{\omega} J(\omega) \omega^{m+n},  \label{eqn:comm_Q_P}
\end{gather}

\section{Derivation of the generating function Eq.~\ref{eqn:expval_gen_func} \label{app:deriv_gen_func}}
In order to evaluate the generating function Eq.~\ref{eqn:def_gen_func}, we start by showing the following operator identity: 
\begin{multline}\label{eqn:key_exponential_property}
    \hat{\rho}_\text{eq}^{B} e^{\kappa \hat{q}_j} e^{\chi \hat{p}_j} = Z^{-1} \Tr[-\beta\hat{H}_B + \frac{x_j}{2} \left[(\kappa\coth\frac{x_j}{2} - \im \chi) \hat{q}_j + (\im \kappa + \chi \coth\frac{x_j}{2}) \hat{p}_j \right] ] \times \\
    \exp[-\frac{x_j \beta (\kappa^2 + \chi^2)}{8 \sinh^2(x_j/2)}] \exp[\frac{\kappa^2+\chi^2}{4} \coth(\frac{x_j}{2})] \exp[\im \frac{\kappa\chi}{2}],
\end{multline}
where $\kappa$ and $\chi$ are arbitrary scalars, $x_j = \hbar\omega_j\beta$, and $Z=\Tr[e^{-\beta \hat{H}_B}]$ is the bath partition function. 

To show Eq.~\ref{eqn:key_exponential_property}, we begin by substituting the ladder operator definitions (Eq.~\ref{eqn:ladder_qp}) for $\hat{q}_j$ and $\hat{p}_j$ into the expression. Using the identity $e^{\hat{A}} e^{\hat{B}} = e^{\hat{A} + \hat{B} + \comm{\hat{A}}{\hat{B}}/2}$ when $\comm{\hat{A}}{\hat{B}}$ is a constant, we obtain
\begin{equation}\label{eqn:xp2ladder}
    e^{\kappa \hat{q}_j} e^{\chi \hat{p}_j} = \exp[\frac{(\kappa - \im \chi)}{\sqrt{2}}\hat{b}_j] \exp[\frac{(\kappa - \im \chi)}{\sqrt{2}}\hat{b}_j] \exp(-\frac{\kappa^2 + \chi^2}{4} + \im\frac{\kappa \chi}{2}).
\end{equation}
Then, we utilize the identity $e^{\hat{A}} e^{\hat{B}} = \exp(\hat{A} + \frac{s}{1 - e^{-s}} \hat{B})$ when $\comm{\hat{A}}{\hat{B}} = s\hat{B}$ consecutively to show that 
\begin{multline}\label{eqn:rho_eq_b}
    \exp[-x_j \left(\hat{b}_j^{\dagger}\hat{b}_j + \frac{1}{2}\right)] \exp[\frac{(\kappa - \im \chi)}{\sqrt{2}}\hat{b}_j] = \\ 
    \exp[-x_j \left(\hat{b}_j^{\dagger}\hat{b}_j + \frac{1}{2}\right) + \frac{x_j}{1 - e^{-x_j}} \frac{(\kappa - \im \chi)}{\sqrt{2}}\hat{b}_j], 
\end{multline}
and 
\begin{multline}\label{eqn:rho_eq_b_bdag_1}
    \exp[-x_j \left(\hat{b}_j^{\dagger}\hat{b}_j + \frac{1}{2}\right) + \frac{x_j}{1 - e^{-x_j}} \frac{(\kappa - \im \chi)}{\sqrt{2}}\hat{b}_j] \exp[\frac{(\kappa + \im \chi)}{\sqrt{2}}\hat{b}_j^{\dagger}] \exp[-\frac{\kappa^2 + \chi^2}{2 (1 - e^{-x_j})}]  = \\
    \exp[-x_j \left(\hat{b}_j^{\dagger}\hat{b}_j + \frac{1}{2}\right) + \frac{x_j}{1 - e^{-x_j}} \frac{(\kappa - \im \chi)}{\sqrt{2}}\hat{b}_j + \frac{-x_j}{1 - e^{x_j}} \frac{(\kappa + \im \chi)}{\sqrt{2}}\hat{b}_j^{\dagger}] \times\\ 
    \exp[- \frac{-x_j}{1 - e^{x_j}} \frac{\kappa^2 + \chi^2}{2 (1 - e^{-x_j})}].
\end{multline}
Finally, we notice that the second and third terms in the first exponential on the right-hand side of Eq.~\ref{eqn:rho_eq_b_bdag_1} follow the pattern $a \hat{X} + (a-1)\hat{X}^{\dagger}$, where $a = (1 - e^{-x_j})^{-1}$ and $\hat{X} = \frac{\kappa - \im\chi}{\sqrt{2}} \hat{b}_j$. Utilizing the identity $a \hat{X} + (a-1)\hat{X}^{\dagger} = \frac{1}{2}(\hat{X} - \hat{X}^{\dagger}) +  (a - \frac{1}{2}) (\hat{X} + \hat{X}^{\dagger})$,  we can re-express Eq.~\ref{eqn:rho_eq_b_bdag_1} in terms of $\hat{q}_j$ and $\hat{p}_j$. Combining Eqs.~\ref{eqn:xp2ladder}, \ref{eqn:rho_eq_b}, and \ref{eqn:rho_eq_b_bdag_1}, we derive the operator identity Eq.~\ref{eqn:key_exponential_property}.

We then directly apply Eq.~\ref{eqn:key_exponential_property} to the generating function Eq.~\ref{eqn:def_gen_func} and have 
\begin{equation}\label{eqn:expo_g}
    f(\{\lambda_{\alpha}\}, \{\lambda_{\gamma}'\}) = \exp[g(\{\lambda_{\alpha}\}, \{\lambda_{\gamma}'\})], \\
\end{equation}
where the exponent is given by
\begin{gather}
    g(\{\lambda_{\alpha}\}, \{\lambda_{\gamma}'\}) = -\beta \hat{H}_B + C_{q} + C_{p} + C_{\alpha} + C_{\gamma} + C_{\alpha \gamma}, \\
    C_{q} = \sum_j \frac{x_j}{2} c_j \left(\coth\frac{x_j}{2}\sum_{\alpha} \lambda_{\alpha} \omega_j^{m_{\alpha}} - \im \sum_{\gamma} \lambda_{\gamma}' \omega_j^{n_{\gamma}} \right) \hat{q}_j, \\
    C_{p} = \sum_j \frac{x_j}{2} c_j \left(\im\sum_{\alpha} \lambda_{\alpha} \omega_j^{m_{\alpha}} +  \coth\frac{x_j}{2} \sum_{\gamma} \lambda_{\gamma}' \omega_j^{n_{\gamma}} \right) \hat{p}_j, \\
    C_{\alpha} = \frac{\hbar\beta}{4\pi} \sum_{\alpha} \sum_{\alpha'} \lambda_{\alpha} \lambda_{\alpha'} \int_{0}^{\infty}\dd{\omega} J(\omega) \omega^{m_{\alpha} + m_{\alpha'}}\left[\frac{2}{\hbar\beta} \coth(\frac{\hbar\omega\beta}{2}) - \frac{\omega}{\sinh^2(\hbar\omega\beta/2)}\right], \\
    C_{\gamma} = \frac{\hbar\beta}{4\pi} \sum_{\gamma} \sum_{\gamma'} \lambda_{\gamma}' \lambda_{\gamma'}' \int_{0}^{\infty}\dd{\omega} J(\omega) \omega^{n_{\gamma} + n_{\gamma'}}\left[\frac{2}{\hbar\beta} \coth(\frac{\hbar\omega\beta}{2}) - \frac{\omega}{\sinh^2(\hbar\omega\beta/2)}\right], \\
    C_{\alpha\gamma} = \frac{\im}{\pi} \sum_{\alpha} \sum_{\gamma} \lambda_{\alpha} \lambda_{\gamma}' \int_{0}^{\infty} \dd{\omega} J(\omega) \omega^{m_{\alpha} + n_{\gamma}}.
\end{gather}
Here, we have utilized the definition of $J(\omega)$ (Eq.~\ref{eqn:spec_den}) to convert the summations over modes $j$ in $C_{\alpha}$, $C_{\gamma}$, and $C_{\alpha\gamma}$ to integrals over $\omega$. To proceed, we complete the squares for $-\beta \hat{H}_B + C_q + C_p$: 
\begin{multline}\label{eqn:complete_sq_q_gen}
    -\frac{x_j}{2}\left[\hat{q}_j^2 - c_j \left(\coth\frac{x_j}{2}\sum_{\alpha} \lambda_{\alpha} \omega_j^{m_{\alpha}} - \im \sum_{\gamma} \lambda_{\gamma}' \omega_j^{n_{\gamma}} \right) \hat{q}_j\right] = \\
    -\frac{x_j}{2}\left[\hat{q}_j - \frac{c_j}{2} \left(\coth\frac{x_j}{2}\sum_{\alpha} \lambda_{\alpha} \omega_j^{m_{\alpha}} - \im \sum_{\gamma} \lambda_{\gamma}' \omega_j^{n_{\gamma}} \right)\right]^2 + \\
    \sum_j \frac{x_j}{8} c_j^2 \left[\coth^2(\frac{x_j}{2}) \sum_{\alpha} \sum_{\alpha'} \lambda_{\alpha} \lambda_{\alpha'}\omega_j^{m_{\alpha} + m_{\alpha'}} - \sum_{\gamma} \sum_{\gamma'} \lambda_{\gamma}' \lambda_{\gamma'}' \omega_j^{n_{\gamma} + n_{\gamma'}} \right] + \\
    \sum_j -\im \frac{x_j}{4} c_j^2  \coth(\frac{x_j}{2}) \sum_{\alpha}\sum_{\gamma} \lambda_{\alpha} \lambda'_{\gamma} \omega_j^{m_{\alpha}+n_{\gamma}}.
\end{multline}
\begin{multline}\label{eqn:complete_sq_p_gen}
    -\frac{x_j}{2}\left[\hat{p}_j^2 - c_j \left(\im\sum_{\alpha} \lambda_{\alpha} \omega_j^{m_{\alpha}} + \coth\frac{x_j}{2} \sum_{\gamma} \lambda_{\gamma}' \omega_j^{n_{\gamma}} \right) \hat{p}_j\right] = \\
    -\frac{x_j}{2}\left[\hat{p}_j - \frac{c_j}{2} \left(\im\sum_{\alpha} \lambda_{\alpha} \omega_j^{m_{\alpha}} + \coth\frac{x_j}{2} \sum_{\gamma} \lambda_{\gamma}' \omega_j^{n_{\gamma}} \right)\right]^2 +\\
    \sum_j \frac{x_j}{8} c_j^2 \left[-\sum_{\alpha} \sum_{\alpha'} \lambda_{\alpha} \lambda_{\alpha'}\omega_j^{m_{\alpha} + m_{\alpha'}} + \coth^2(\frac{x_j}{2})  \sum_{\gamma} \sum_{\gamma'} \lambda_{\gamma}' \lambda_{\gamma'}' \omega_j^{n_{\gamma} + n_{\gamma'}} \right] + \\
    \sum_j +\im \frac{x_j}{4} c_j^2  \coth(\frac{x_j}{2}) \sum_{\alpha}\sum_{\gamma} \lambda_{\alpha} \lambda'_{\gamma} \omega_j^{m_{\alpha}+n_{\gamma}}.
\end{multline}
Using Eqs.~\ref{eqn:complete_sq_q_gen} and \ref{eqn:complete_sq_p_gen}, we rewrite $g(\{\lambda_{\alpha}\}, \{\lambda_{\gamma}'\})$ as the Boltzmann exponent of a shifted harmonic oscillator, plus additional constant terms that depend on coefficients $\{\lambda_\alpha\}$ and $\{\lambda_{\gamma}'\}$. Since shifting a harmonic oscillator does not affect its partition function, only these constant terms contribute to the final expression for $f(\{\lambda_{\alpha}\}, \{\lambda_{\gamma}'\})$. Finally, by combining Eqs.~\ref{eqn:expo_g}-\ref{eqn:complete_sq_p_gen}, we obtain the generating function expression given in Eq.~\ref{eqn:expval_gen_func} of the main text.

\section{\label{app:parameters}Robustness of MKCT across parameter regimes of linearly coupled spin-boson models}
In the main text, we have demonstrated the efficacy of MKCT in a parameter regime relevant to molecular spectroscopy, where $\Delta \gg \omega_D > \beta^{-1}$. Assuming room temperature thermal energy, FIGs.~\ref{fig:linear_sb_time} and~\ref{fig:linear_sb_freq} correspond to an absorption spectrum with a splitting of approximately 2.5 eV, where both low-frequency ($\hbar\omega \leq \beta^{-1}$) and high-frequency ($\hbar\omega > \beta^{-1}$) phonons contribute to the dissipative dynamics. While this regime is representative of many molecular systems, it is important to assess the broader applicability of MKCT across other dynamical conditions. To this end, we provide additional benchmarks in the following, where the robustness of MKCT is systematically tested against a wide range of spin-boson model parameters, including varying bias and system timescales, bath timescales, coupling strengths, and temperatures.

\begin{figure}[htbp]
    \centering
    \includegraphics[width=0.5\linewidth]{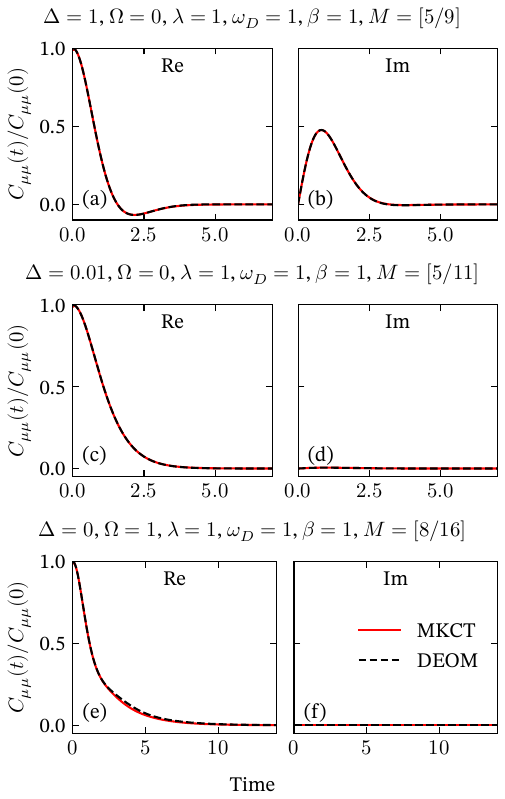}
    \caption{Benchmark of MKCT with varying system parameters in the spin-boson model. Panels (a, b) show the reference scenario with comparable time scales between the system and bath. Panels (c, d) illustrate a case with minimal bias and slow system dynamics, while panels (e, f) present results for a completely unbiased system. The spin-boson model parameters, as well as the Pad\'{e} approximant orders $[M_1/M_2]$, are indicated above each panel. The MKCT cutoff is consistently set to $n=1$. The linear coupling strength is set to $\alpha_1 = 1$.}
    \label{fig:sys_param_roubust}
\end{figure}

\begin{figure}[htbp]
    \centering
    \includegraphics[width=0.5\linewidth]{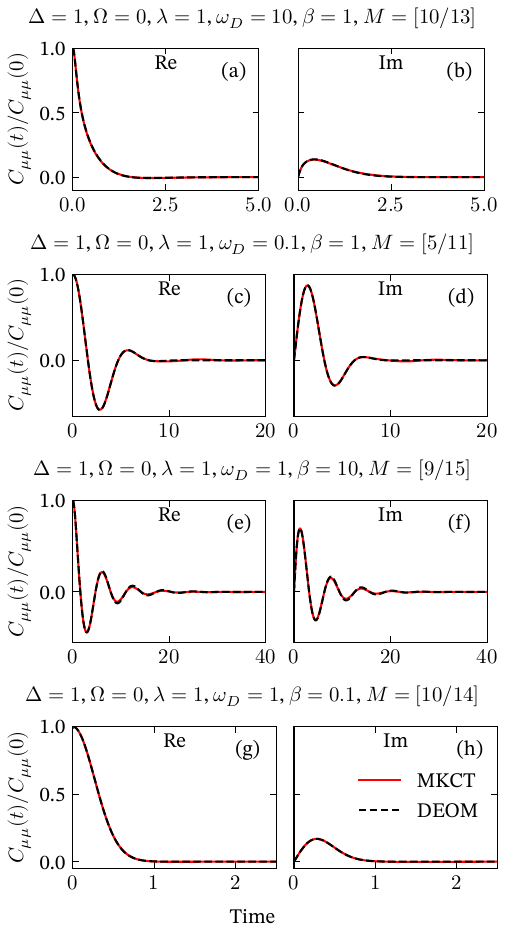}
    \caption{Benchmark of MKCT with varying bath parameters in the spin-boson model. Panels (a, b) and (c, d) demonstrate cases with fast and slow bath dynamics, respectively. Panels (e, f) and (g, h) correspond to low and high temperatures. Parameter annotations and the MKCT cutoff $n$ follow those used in Fig.~\ref{fig:sys_param_roubust}.}
    \label{fig:bath_param_roubust}
\end{figure}

\begin{figure}[htbp]
    \centering
    \includegraphics[width=0.5\linewidth]{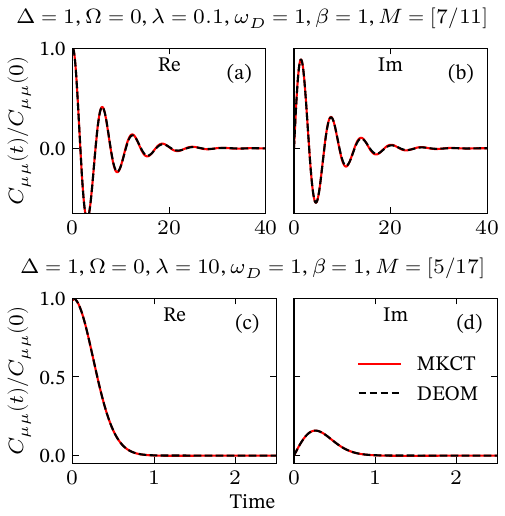}
    \caption{Benchmark of MKCT with varying system-bath coupling strenght in the spin-boson model. Panels (a, b) and (c, d) demonstrate cases with strong and weak bath dynamics, respectively. Parameter annotations and the MKCT cutoff $n$ follow those used in Fig.~\ref{fig:sys_param_roubust}.}
    \label{fig:cp_param_roubust}
\end{figure}

To demonstrate the robustness of MKCT, we benchmarked it across a range of system parameters in the spin-boson model. As shown in FIG.\ref{fig:sys_param_roubust}(a, b), MKCT accurately reproduces DEOM results under balanced conditions ($\Delta = \omega_D = \beta^{-1} = 1$, $\lambda = 1$). This agreement holds under slower system dynamics [FIG.\ref{fig:sys_param_roubust}(c, d)] and in the unbiased limit [FIG.\ref{fig:sys_param_roubust}(e, f)]. FIG.\ref{fig:bath_param_roubust} further confirms that MKCT remains reliable across both fast [panels (a, b)] and slow [panels (c, d)] bath regimes, as well as at low [panels (e, f)] and high [panels (g, h)] temperatures. Finally, FIG.~\ref{fig:cp_param_roubust} demonstrates consistent performance across both weak and strong coupling strengths.

\bibliography{ref}

\end{document}